\documentclass[journal]{IEEEtran}

\usepackage{cite}
\usepackage{latexsym}
\usepackage{graphicx}
\usepackage{amsfonts,amssymb,amsmath}
\usepackage{nccmath}
\usepackage{optidef}
\usepackage{hyperref}
\usepackage{color,soul}
\usepackage{array}
\usepackage[T1]{fontenc}
\usepackage[utf8]{inputenc}
\usepackage{algorithm}
\usepackage{algorithmic}
\usepackage{caption}

\DeclareMathAlphabet{\pazocal}{OMS}{zplm}{m}{n}

\newcommand{\Ub}{\pazocal{U}}

\begin{document}
\title{Bounds on Power and Common Message Fractions for RSMA with Imperfect SIC}
\author{Garima~Chopra,~\IEEEmembership{Member,~IEEE,}
        {Akhileswar Chowdary,~\IEEEmembership{Graduate Student Member,~IEEE,}}
       ~Abhinav~Kumar,~\IEEEmembership{Senior~Member,~IEEE}
        and~Marwa~Chafii,~\IEEEmembership{Member,~IEEE} % <-this % stops a space
\thanks{Garima Chopra is with Chitkara University Institute of Engineering and Technology, Chitkara University, Punjab, India-173212, Akhileswar Chowdary is with NYU WIRELESS, New York University Tandon School of Engineering, Brooklyn, New York, Abhinav Kumar is with the Department of Electrical Engineering, Indian Institute of Technology Hyderabad, Telangana, India-502285, and Marwa Chafii is with the Engineering Division, New York University (NYU) Abu Dhabi, Abu Dhabi and NYU WIRELESS, NYU Tandon School of Engineering, Brooklyn, New York (Email: garima.chopra@chitkara.edu.in, akhileswar.chowdary@nyu.edu, abhinavkumar@ee.iith.ac.in, marwa.chafii@nyu.edu). This work is supported in part by the SERB NPDF--project: PDF/2020/001251 and DST NMICPS through TiHAN Faculty fellowship of Dr. Abhinav Kumar.}% <-this % stops a space
}
\markboth{Journal of \LaTeX\ Class Files,~Vol.~14, No.~8, August~2015}%
{Shell \MakeLowercase{\textit{et al.}}: Bare Demo of IEEEtran.cls for IEEE Journals}
\maketitle
\begin{abstract}
Rate-Splitting multiple access (RSMA) has emerged as a key enabler in improving the performance of the beyond fifth-generation (5G) cellular networks. The existing literature has typically considered the sum rate of the users to evaluate the performance of RSMA. However, it has been shown in the existing works that maximizing the sum rate can result in asymmetric user performance. It significantly enhances one user's rate at the cost of the rate of another RSMA user. Further, imperfections can reduce the performance of successive interference cancellation (SIC)-based RSMA. Therefore, in this letter, we consider the imperfection in SIC and derive suitable bounds on fractions of the power allocated for common and private messages and the fraction of common message intended for each user in an RSMA pair such that their individual RSMA rates are greater than their respective orthogonal multiple access (OMA) rates. Through simulations, we validate the derived bounds. We show that they can be used to appropriately select the RSMA parameters resulting in users' RSMA rates being better than their respective OMA rates.
\end{abstract}
\begin{IEEEkeywords}
Beyond Fifth Generation (5G) cellular networks, Rate-Splitting Multiple Access (RSMA), Successive Interference Cancellation (SIC).
\end{IEEEkeywords}
\IEEEpeerreviewmaketitle
\section{Introduction}
\IEEEPARstart{R}{ate} splitting multiple access (RSMA) is gaining popularity due to its robustness, reliability, and high throughput in comparison to other multiple access schemes: orthogonal multiple access (OMA), non-orthogonal multiple access (NOMA) \cite{l1}. RSMA is an established multiple access scheme for which NOMA and space division multiple access (SDMA) act as special cases. For a downlink RSMA, the user's message is split into common and private messages at the transmitter, and the successive interference cancellation (SIC) is executed at the receiver's side \cite{l1}. 

In \cite{l8}, the authors have worked on the sum-rate maximization for wireless networks that use RSMA in the downlink. It has been shown that RSMA can outperform orthogonal frequency division multiple access (OFDMA) and NOMA in terms of data rate with $23.5 \%$ and $19.6 \%$ gains, respectively. The authors in \cite{l4} have developed a novel transmission scheme that utilizes RSMA with artificial noise and adaptive beamforming to maximize the secrecy sum rate. In \cite{l3}, the authors have applied RSMA for a multi-group and multi-cast scenario where each message is intended for a group of users. They have also designed the physical layer and performed a link-level simulation for RSMA and SDMA by assuming cellular and multi-beam satellite systems. The comparative analysis in \cite{l5} shows that RSMA is more robust and efficient than NOMA for multiple-input multiple-output (MIMO) settings. An extensive study on key multiple access technologies for aerial networks has been presented in \cite{l6}. The authors in \cite{l6} have modeled and analyzed the weighted sum-rate performance of two user networks served by an RSMA-based aerial base station (BS). The authors in \cite{l7} have derived an optimal power allocation strategy and demonstrated that RSMA with the proposed strategy is robust to degrading performance due to user mobility compared to the conventional MIMO strategies.

Most of the existing works on RSMA have concentrated on the sum-rate maximization and other aspects related to the sum rate of the users. However, a study on the individual user rates in RSMA is required. \textit{An asymmetric increase in user rates can result in unfairness between RSMA users}. Motivated by this, we derive the bounds on the power allocation of common and private messages and the fraction of the individual common message for which the RSMA rates of strong and weak users will be greater than their respective OMA rates simultaneously. To the best of our knowledge, this is the first work that derives the bounds on power allocation and common message fraction by considering the imperfection in SIC for RSMA. Therefore, \textit{by fixing the power allocation fractions of common and their individual private messages and the fraction of common message within the derived bounds, individual rates of RSMA users will always be greater than their respective OMA rates}. The main contributions of the letter are as follows.
\begin{itemize}
    \item We derive bounds on power allocation coefficients of common and private messages and the fraction of individual common messages for RSMA for a two-user scenario with imperfect SIC.
    \item Through extensive simulations, we show that the bounds are close to the numerically obtained values.
\end{itemize}
The organization of the letter is as follows. In Section \ref{system_model}, the system model is presented. The bounds on the fraction of common message for a user and power allocation coefficients for common and private messages are derived in Section \ref{bounds}. The derived bounds are validated in Section \ref{results}. The conclusion and future work are presented in Section \ref{conclusion}.  

% In the downlink RSMA, the message intended for a users ($msg_1$) and ($msg_2$) is divided into corresponding common and private sub-messages (i.e. $msg^c_i$ and $msg^p_i$), where $i$ is user index). The common part of all the users are combined together to form a single common stream using linear precoder before transmitting to users, and distinct private streams. The resultant streams are then superimposed together into one single stream by the transmitter. The receiver tries to decode the common sub-message from the received superimposed signal then using SIC, the private sub-message is then decoded. RSMA undergoes performance limitations. Firstly, the decoding order plays a major role in determining the performance at the receiver side. Secondly, in the downlink, each and every user needs to perform SIC to extract its own private message at least once. The number of times the SIC to be performed by the each user depends on two factors: 1) order of its channel gain in the pair, and 2) number of users in a RSMA pair. Despite of receiver complexity, RSMA initiates flexibility to partially treat interference as noise and partially decode interference. In \cite{1}, the comprehensive survey on multiple access schemes has been presented and also discussed the research challenges. The multi-carrier resource allocation problem has been investigated in \cite{2}, to maximize the sum-rate of the system by formulating model of joint-power distribution and sub-carrier matching. 

\section{System Model} \label{system_model}
\begin{figure}[t]
    \centering
    \captionsetup{justification=centering}
    \includegraphics[width=9cm,height=10cm,keepaspectratio]{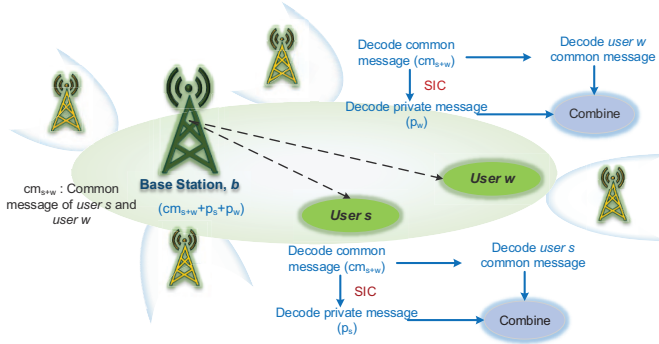}
    \vspace{-0.21in}
    \caption{System Model.}\vspace{-0.2in}
    \label{fig:system_model}
\end{figure}

We consider a downlink RSMA system for a BS as depicted in Fig. \ref{fig:system_model}. Let $\Ub = \lbrace 1,2,3,...,N \rbrace$ be the set of users associated with the BS $b$ under consideration for a given user association scheme and $N$ be the total number of users connected with BS $b$. In downlink transmission, the signal-to-interference-plus-noise ratio (SINR) from BS $b$ to user $u$ ($u \in \Ub$) on a subchannel $m$ in case of OMA is given as
\begin{equation} \label{1}
    \gamma_u=\frac{P_t \|h_u\|^2}{\sigma^2+I_u},
\end{equation}
where $P_t$ is the maximum transmit power of BS $b$, $\|h_u\|^2$ is the channel gain between user $u$ and BS $b$, and $\sigma^2$ is the noise variance. $I_u= \sum_{\hat{b} \in \mathcal{B}, \hat{b} \neq b} P_t^{\hat{b}} \|h_u\|^2$ is the total interference from neighboring BSs, where $P_t^{\hat{b}}$ and $\mathcal{B}$ is the transmit power of neighboring BSs and set of BSs in a given area, respectively. The downlink rate for user $u$ based on the log-rate model is given as \cite{l1}
\begin{equation} \label{2}
    R^{\text{oma}}_{u}=\frac{1}{2} \log_2(1+\gamma_u),
\end{equation}

We consider a 1-layer RSMA system for a pair of RSMA users, i.e., $s$ and $w$ $(s\;\text{and}\;w \in \Ub)$. The proposed bounds can be implemented for all the RSMA pairs. Without loss of generality, we assume that $\|h_{s}\|^2 > \|h_{w}\|^2$. Therefore, we denote $s$ as the strong user and $w$ as the weak user. The users' messages, $s$ and $w$ are divided into their respective common and private messages. Using RSMA, different power levels are assigned to common and private messages of strong and weak users. Let $P_c$, $P_{ps}$, and $P_{pw}$ be the power of common stream of $s$ and $w$, private stream of $s$, and private stream of $w$, respectively, such that $P_c+P_{ps}+P_{pw} \leq P_t$.

The receiver of the respective users performs SIC to decode their messages. At the receiver side, the user $s$ decodes the common message by treating the interference from the private messages as noise. Then, the user $s$ subtracts the common message from the received signal to extract its own private message by treating other users' private messages as noise. Therefore, the SINR of the common message of user $s$ is given as follows.
\begin{equation}\label{3a}
    \gamma_{cs} = \frac{P_c \|h_s\|^2}{(P_{ps}+P_{pw}) \|h_s\|^2 + \sigma^2 + I_u}.
\end{equation}
Dividing numerator and denominator of (\ref{3a}) by $(\sigma^2+I_u)$ and replacing $\|h_s\|^2/ (\sigma^2+I_u)$ with $\gamma_s/P_t$ from (\ref{1}), we get
\begin{equation} \label{3b}
    \gamma_{cs} = \frac{P_c \gamma_{s}}{(P_t-P_c)\gamma_{s} + P_t}.
\end{equation}
Additionally, the SINR of the private message of user $s$ is
\begin{equation} \label{4}
    \gamma_{ps}=\frac{P_{ps} \gamma_{s}}{\beta P_c \gamma_s+P_{pw}\gamma_{s} + P_t},
\end{equation}
where $P_c=\alpha_c P_t$, $P_{ps}=\lambda (P_t-P_c)$, $P_{pw}=(1-\lambda) (P_t-P_c)$, $\alpha_c \in (0,1)$ is the power fraction of common message, $\lambda \in (0,1)$ is the power fraction of the private message of $s$, and $\beta \in [0,1]$ represents the coefficient of SIC imperfection. Similarly, for user $w$, the SINRs of common and private messages are, respectively,
\begin{equation} \label{5}
    \gamma_{cw}=\frac{P_c \gamma_w}{(P_t-P_c)\gamma_w + P_t},
\end{equation}
\begin{equation} \label{6}
    \gamma_{pw}=\frac{P_{pw} \gamma_{w}}{\beta P_c \gamma_{w} + P_{ps} \gamma_w +P_t}.
\end{equation}
Therefore, the RSMA rates for user $s$ and $w$ based on the SINRs obtained in (\ref{3b}), (\ref{4}), (\ref{5}), and (\ref{6}) are
\begin{equation} \label{7}
    R^{\text{rsma}}_{s}=\tau R^{\text{comm.}}+R^{\text{priv.}}_{s}, 
\end{equation}
\begin{equation} \label{8}
        R^{\text{rsma}}_{w}=(1-\tau) R^{\text{comm.}}+R^{\text{priv.}}_{w},
\end{equation}
where $R^{\text{comm.}}_{x}=\log_2(1+\gamma_{cx})$, $R^{\text{priv.}}_{x}=\log_2(1+\gamma_{px})$, $x \in \lbrace s,w \rbrace$, and $R^{\text{comm.}}=\min(R^{\text{comm.}}_{s},R^{\text{comm.}}_{w})$ \cite{l1}. $\tau \in (0,1)$ is the fraction of common message intended only for user $s$ from $R^{\text{comm.}}$. From the expression of $R^{\text{comm.}}$, the common rate relies on the common message rate of the weak user, i.e., $w$ \cite{l1}. Hence, we take $R^{\text{comm.}} \approx R^{\text{comm.}}_{w}$ for evaluating the bounds on the power fractions in the following section. The sum-rate for an RSMA pair is computed using (\ref{7}) and (\ref{8}) as
\begin{equation} \label{9}
    SR^{\text{rsma}}=R^{\text{comm.}}+R^{\text{priv.}}_{s}+R^{\text{priv.}}_{w}.
\end{equation}
Using (\ref{2}), the OMA sum-rate for the same pair is given as
\begin{equation} \label{10}
    SR^{\text{oma}}=R^{\text{oma}}_{s}+R^{\text{oma}}_{w}.
\end{equation}

\section{Bounds on Common Message fraction of $s$ ($\tau$), Power Fraction of Common Message ($\alpha_c$), and Power Fraction of Private Message of $s$ ($\lambda$)} \label{bounds}
In this section, we derive bounds on $\tau$, $\alpha_c$, and $\lambda$ for which the individual RSMA rates of  user $s$ and $w$ ($R^{\text{rsma}}_{s}$ and $R^{\text{rsma}}_{w}$) should be greater than their respective OMA rates ($R^{\text{oma}}_{s}$ and $R^{\text{oma}}_{w}$).
\subsection{Bounds on $\tau$}
We consider that $R^{\text{rsma}}_{s}>R^{\text{oma}}_{s}$ and $R^{\text{rsma}}_{w}>R^{\text{oma}}_{w}$ for user $s$ and $w$, respectively. Using these constraints and (\ref{2}), (\ref{7}), (\ref{8}), we derive the upper and lower bounds on $\tau$. The constraints are expressed as follows
\begin{equation} \label{tau1}
    \tau R^{\text{comm.}} + R^{\text{priv.}}_{s} > R^{\text{oma}}_{s}.
\end{equation}
Substituting $R^{\text{comm.}}$, $R^{\text{priv.}}_{s}$, and $R^{\text{oma}}_{s}$ with their respective expressions from (\ref{5}), (\ref{4}), and (\ref{2}) in (\ref{tau1}), we obtain
\begin{equation} \label{tau2}
    \tau > \frac{1}{\log_2(1+\gamma_{cw})} \bigg[\frac{1}{2} \log_2(1+\gamma_s) - \log_2(1+\gamma_{ps}) \bigg].
\end{equation}
Substituting $\gamma_{cw}$ and $\gamma_{ps}$ from (\ref{5}) and (\ref{4}) in (\ref{tau2}) and further solving (\ref{tau2}), we get
\begin{equation} \label{tau_lb}
    \tau > \frac{\log_2\bigg(\frac{\sqrt{(1+\gamma_s)}(\beta \alpha_c \gamma_s+ \gamma_{s} (1-\lambda) (1 - \alpha_c) + 1)}{\alpha_c \gamma_s (\beta - 1) + (\gamma_s + 1) }\bigg)}{\log_2\big(\frac{(\gamma_w + 1)}{\gamma_w (1-\alpha_c) + 1}\big)}.
    %\log_2\bigg(\frac{(\beta - 1)P_c \gamma_s + (\gamma_s + 1) P_t}{\beta P_c \gamma_s+ (1-\lambda) (P_t - P_c) \gamma_{s} + P_t}\bigg) \bigg],
\end{equation}
Similarly, the constraint for user $w$ is given as follows
\begin{equation} \label{tau4}
    (1-\tau) R^{\text{comm.}} + R^{\text{priv.}}_{w} > R^{\text{oma}}_{w}.
\end{equation}
Using $\gamma_{cw}$ and $\gamma_{pw}$ from (\ref{5}) and (\ref{6}) in (\ref{tau4}), we get
\begin{equation} \label{tau_ub}
    \hspace{-0.1in}\tau < \frac{\log_2\bigg(\frac{(\gamma_w + 1)(\alpha_c\gamma_w(\beta -1) + (\gamma_w + 1))}{(\sqrt{1+\gamma_w})(\gamma_w (1 - \alpha_c) + 1)(\beta \alpha_c \gamma_w + \lambda \gamma_w (1 - \alpha_c) + 1)}\bigg)}{\log_2\bigg(\frac{(\gamma_w + 1)}{\gamma_w (1 - \alpha_c) + 1}\bigg)}.
\end{equation}
The bounds derived in (\ref{tau_lb}) and (\ref{tau_ub}), respectively, are considered as lower and upper bound for $\tau$ and referred to as $\tau_{\text{lower}}$ and $\tau_{\text{upper}}$, respectively. Note that the upper and lower bounds on $\tau$ depend on $\alpha_c$, $\beta$, and $\lambda$.

\subsection{Bounds on $\alpha_c$}
The range of $\tau$ is $(0,1)$. Rearranging (\ref{tau1}), we get
\begin{equation}\label{tau_primary}
    \tau > \frac{R^{\text{oma}}_{s} - R^{\text{priv.}}_{s}}{R^{\text{comm.}}}.
\end{equation}
Considering $\tau > 0$ in (\ref{tau_primary}), we obtain,
\begin{equation*}
    \frac{R^{\text{oma}}_{s} - R^{\text{priv.}}_{s}}{R^{\text{comm.}}} > 0 \implies R^{\text{oma}}_{s} > R^{\text{priv.}}_{s},
\end{equation*}
\begin{equation}
    \frac{1}{2}\log_2(1 + \gamma_s) > \log_2(1 + \gamma_{ps}) \implies \gamma_{ps} < \sqrt{1 + \gamma_s} - 1.\label{pc1}
\end{equation}
Substituting the expression of $\gamma_{ps}$ from (\ref{4}) in (\ref{pc1}), and solving further, we get
\begin{equation} \label{pc_lower}
   \alpha_c > \frac{\sqrt{1 + \gamma_s}(\lambda \gamma_s - \gamma_s - 1) + \gamma_s +1}{\gamma_s (\sqrt{1+\gamma_s}(\beta - 1 + \lambda) + 1 - \beta)} \triangleq \alpha_{\text{LB}}.
\end{equation}
The $\alpha_c$ should satisfy (\ref{pc_lower}) for a given $\lambda$ and $\beta$ for $\tau_{\text{lower}}$ to be greater than 0. Similarly, considering and rearranging (\ref{tau4}), we get,
\begin{equation} \label{pc2}
    \tau < 1 - \bigg(\frac{R^{\text{oma}}_{w} - R^{\text{priv.}}_{w}}{R^{\text{comm.}}}\bigg).
\end{equation}
% Applying the constraint that RHS of the inequality in (\ref{pc2}) should be greater than 0, and after solving, we get the following inequality
% \begin{multline}\label{pc2_1}
%     \alpha_c^2\gamma_w^2\Big[\lambda-\beta\Big] + \alpha_c\gamma_w\Big[\beta(\gamma_w+1)-\lambda(2\gamma_w+1)\\ -\frac{(\gamma_w+1)(\beta-1)}{\sqrt{1+\gamma_w}}-1\Big]+\Big[\gamma_w(\lambda(1+\gamma_w) + 1)\\ - \frac{(\gamma_w + 1)^2}{\sqrt{1+\gamma_w}}+1\Big] < 0
% \end{multline}
% The $\alpha_c$ should also satisfy (\ref{pc2_1}) for a given $\lambda$ and $\beta$.  
Solving (\ref{pc2}) by imposing the constraint that the right-hand side (RHS) of the expression should be less than 1, we get,
\begin{equation*}
    \frac{R^{\text{oma}}_{w} - R^{\text{priv.}}_{w}}{R^{\text{comm.}}} > 0 \implies R^{\text{oma}}_{w} > R^{\text{priv.}}_{w},
\end{equation*}
\begin{equation}
    \gamma_{pw} < \sqrt{1+\gamma_w} - 1. \label{pc3}
\end{equation}
Replacing $\gamma_{pw}$ with (\ref{6}) in (\ref{pc3}) and further solving, we get
\begin{equation} \label{pc_upper}
    \alpha_c < \frac{\sqrt{1 + \gamma_w} (\lambda \gamma_w + 1)-(1 + \gamma_w)}{\gamma_w(\sqrt{1 + \gamma_w}(\lambda-\beta) + \beta - 1)}.
\end{equation}
Along with (\ref{pc_lower}), the $\alpha_c$ should also satisfy (\ref{pc_upper}) for a given $\lambda$ and $\beta$ for $\tau_{\text{upper}}$ to be less than 1. We also need to select an $\alpha_c$ such that $\tau_{\text{lower}} < \tau_{\text{upper}}$. Hence, solving the inequality $(\ref{tau_lb})<(\ref{tau_ub})$, we get the following cubic equation in $\alpha_c$.
\begin{multline} \label{pc2_2}
    -\alpha_c^3\gamma_s\gamma_w^2 B + \alpha_c^2\Bigg[\gamma_w \bigg(\gamma_s B(\gamma_w + 1)-C\bigg)\\- \frac{\gamma_s\gamma_w(\beta-1)^2(\gamma_w+1)}{A}\Bigg] + \alpha_c\Bigg[\gamma_w(C-D)+C -\\ \frac{(\beta-1)(\gamma_w+1)(\gamma_s-\gamma_w)}{A}\Bigg] + \Bigg[(\gamma_w+1)\\\left(D-\frac{(\gamma_s+1)(\gamma_w+1)}{A}\right) \Bigg] < 0,
\end{multline}
where
\begin{align}
    A &= \sqrt{1+\gamma_s}\sqrt{1+\gamma_w},\nonumber\\
    B &= \beta(\beta-1) + \lambda(1-\lambda),\nonumber\\
    C &= \beta(\gamma_s+\gamma_w) + \gamma_s\gamma_w(\beta-\lambda(1-\lambda))\nonumber \\ &\;\;\;\;\;\;\;\;\;\;\;\;\;\;\;\;- \lambda(\gamma_s\gamma_w(1-\lambda)+\gamma_w-\gamma_s) - \gamma_s, \nonumber\\
    D &= \lambda(\gamma_s\gamma_w(1-\lambda)+\gamma_w-\gamma_s)+\gamma_s+1. \nonumber
\end{align}
The $\alpha_c$ should satisfy (\ref{pc_lower}), (\ref{pc_upper}), and (\ref{pc2_2}) for a given $\lambda$ and $\beta$. The expressions in (\ref{pc_lower}), (\ref{pc_upper}), and (\ref{pc2_2}) are functions of $\lambda$ and $\beta$. Thus, $\lambda$ and $\beta$ have a significant impact on the value of $\alpha_c$. Therefore, we need to choose a $\lambda$ such that the bounds of $\alpha_c$ always lie between 0 and 1. Considering this condition, we next derive bounds on $\lambda$. 

\subsection{Bounds on $\lambda$}\label{lambda bounds}
We consider the constraint that $\alpha_c$ lies between $(0,1)$. Applying this constraint on the bounds of $\alpha_c$ derived in (\ref{pc_lower}) and (\ref{pc_upper}), we derive the bounds on $\lambda$. Considering numerator of (\ref{pc_lower}) to be greater than $0$, and further solving, we get
\begin{equation} \label{lam1}
    \lambda > \frac{(\sqrt{1+\gamma_s}-1)\sqrt{1+\gamma_s}}{\gamma_s}.
\end{equation}
Now, applying the positivity constraint on the denominator of (\ref{pc_lower}), we obtain,
\begin{equation} \label{lam2}
    \lambda > \frac{(\sqrt{1+\gamma_s}-1)(1-\beta)}{\sqrt{1 + \gamma_s}}.
\end{equation}
By applying the constraint on (\ref{pc_lower}) that $\alpha_c<1$, we get
\begin{equation} \label{lam3}
    \frac{\sqrt{1 + \gamma_s}(\lambda \gamma_s - \gamma_s - 1) + \gamma_s +1}{\gamma_s (\sqrt{1+\gamma_s}(\beta - 1 + \lambda) + 1 - \beta)} < 1,
\end{equation}
Considering that (\ref{lam2}) is true, we obtain,
\begin{equation} \label{lam7}
    \beta > \frac{-1}{\gamma_s}.
\end{equation}
We already know that $\beta \in [0,1]$. Hence, $\beta$ always satisfies (\ref{lam7}). This implies that the lower bound on $\alpha_c$ in (\ref{pc_lower}) is always less than 1.
Similarly, applying the positivity constraint on the numerator and denominator of (\ref{pc_upper}), we get
\begin{figure}[t]
     \centering
     \captionsetup{justification=raggedright,singlelinecheck=false}
     \includegraphics[width=8cm,height=10cm,keepaspectratio]{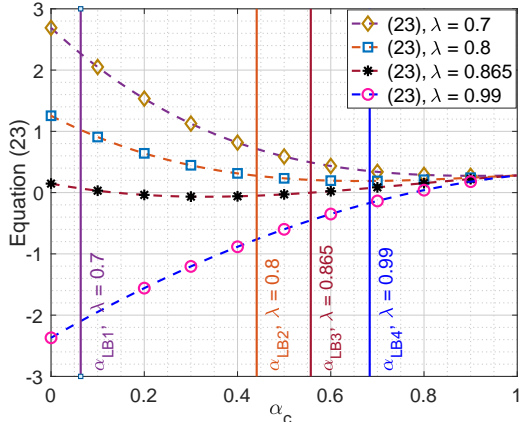}
     \vspace{-0.1in}
     \caption{Variation of the cubic equation and region of operation of $\alpha_c$ with respect to (w.r.t.) $\lambda$.}\vspace{-0.2in}
     \label{fig:fig2}
\end{figure}
\begin{figure}[t]
     \centering
     \captionsetup{justification=raggedright,singlelinecheck=false}
     \includegraphics[width=8cm,height=10cm,keepaspectratio]{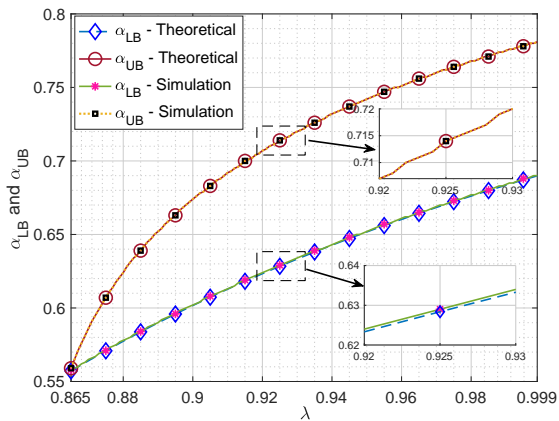}
     \vspace{-0.1in}
     \caption{Variation of the suitable upper bound (\ref{pc2_2}) and lower bounds of $\alpha_c$ (\ref{pc_lower}) w.r.t. $\lambda$.}\vspace{-0.2in}
     \label{fig:fig3}
\end{figure}
\begin{align}
    \lambda &> \frac{\sqrt{1+\gamma_w} -1}{\gamma_w},\label{lam4}\\
    \lambda &> \frac{\beta(\sqrt{1+\gamma_w}-1)+1}{\sqrt{1+\gamma_w}}.\label{lam5}
\end{align}
By setting (\ref{pc_upper}) less than $1$ and assuming (\ref{lam5}) to be true, we get
\begin{equation} \label{lam6}
    \beta < \frac{-1}{\gamma_w}.
\end{equation}
The condition obtained in (\ref{lam6}) cannot be true as $\beta \in [0,1]$. Hence, the upper bound on $\alpha_c$ in (\ref{pc_upper}) cannot be less than $1$. Therefore, it is not a strict upper bound on $\alpha_c$. Further, we have also derived a few other bounds on $\lambda$ by applying the constraint that the numerator and denominator of (\ref{pc_lower}) and (\ref{pc_upper}) are less than 0. However, we found that those bounds are insignificant. Due to space constraints, we are not presenting them in this letter. \vspace{-0.1in}

\subsection{Strict lower bound on $\lambda$ and Strict upper bound on $\alpha_c$}
We have plotted the left-hand side (LHS) of (\ref{pc2_2}) for different values of $\lambda$ and $\alpha_c$ in Fig. \ref{fig:fig2} for $\gamma_s = 6$ dB and $\gamma_w = 2$ dB. Using these SINR values, we have computed the lower bound of $\lambda$ using (\ref{lam1}), (\ref{lam2}), (\ref{lam4}), and (\ref{lam5}) which is approximately equal to $0.7$. Hence, we have plotted the LHS of (\ref{pc2_2}) for various $\lambda$ starting from $0.7$. The values of $\alpha_{\text{LB}}$ for different $\lambda$ are evaluated using (\ref{pc_lower}). The vertical lines in Fig. \ref{fig:fig2} represent the lower bounds of $\alpha_c$ for different $\lambda$. The inequality shown in (\ref{pc2_2}) states that the cubic expression in $\alpha_c$ on the LHS should be less than $0$. Therefore, for a particular $\lambda$, we need to select $\alpha_c$ such that it satisfies $\alpha_c > \alpha_{\text{LB}}$ and the inequality mentioned in (\ref{pc2_2}). From Fig. \ref{fig:fig2}, it is observed that for the lower bound on $\lambda$ computed using (\ref{lam1}), (\ref{lam2}), (\ref{lam4}), and (\ref{lam5}) no $\alpha_c$ satisfies (\ref{pc2_2}). As we increase $\lambda$, we observe that for $\lambda \geq 0.865$ there exists $\alpha_c$ which satisfies $\alpha_c > \alpha_{\text{LB}}$ and (\ref{pc2_2}). Therefore, the lower bound of $\lambda$ obtained using Section \ref{lambda bounds} is not a strict lower bound. The suitable lower bound of $\lambda$ is the lowest value of $\lambda$ for which there exists a solution of (\ref{pc2_2}) which is denoted as $\lambda_{\text{lower}}$. In addition, (\ref{pc2_2}) provides the suitable upper bound of $\alpha_c$ denoted by $\alpha_{\text{UB}}$ which is the point where the curve cuts the x-axis after the vertical line. Similarly, we have plotted the $\alpha_{\text{LB}}$ and $\alpha_{\text{UB}}$ for $\lambda \geq 0.865$ (obtained from Fig. \ref{fig:fig2}). Since the bounds of $\alpha_c$ depend on $\lambda$ as shown in (\ref{pc_lower}) and (\ref{pc2_2}), we have plotted variation of $\alpha_c$ w.r.t $\lambda \geq\lambda_{LB}$ in Fig. \ref{fig:fig3}. It is observed that as the value of $\lambda$ increases, the region of operation of $\alpha_c$ widens. Summarizing Fig. \ref{fig:fig2} and Fig. \ref{fig:fig3} for a given $\gamma_s$ and $\gamma_w$, there exists a strict lower bound on $\lambda$ and upper bound on $\alpha_c$ whose closed-form expression can be derived using (\ref{pc2_2}). We have further validated these bounds in the following section. The computation of the RSMA rates of the strong and weak users using the derived bounds is presented in Algorithm 1.
\begin{algorithm}[t]\label{alg:algorithm_2}
\caption{Steps for calculating RSMA rates of strong and weak users using the derived bounds.}
\begin{algorithmic}[1]
\STATE INPUTS: $\gamma_s$ and $\gamma_w$.
\STATE OUTPUTS: $\tau$, $\alpha_c$, $\lambda$, $R^{\text{rsma}}_{s}$, and $R^{\text{rsma}}_{w}$ .
\STATE For a given $\gamma_s$ and $\gamma_w$, compute $\lambda_{\text{lower}}$ using (\ref{pc2_2}).
\STATE Select $\lambda > \lambda_{\text{lower}}$.
\STATE Using the selected $\lambda$, compute $\alpha_{\text{LB}}$ and $\alpha_{\text{UB}}$ using (\ref{pc_lower}) and (\ref{pc2_2}), respectively.
\STATE Select an $\alpha_c$ such that $\alpha_{\text{LB}} < \alpha_c < \alpha_{\text{UB}}$.
\STATE Using the selected $\alpha_c$ and $\lambda$, compute $\tau_{\text{lower}}$ and $\tau_{\text{upper}}$ using (\ref{tau_lb}) and (\ref{tau_ub}), respectively.
\STATE Select a $\tau$ such that $\tau_{\text{lower}} < \tau < \tau_{\text{upper}}$.
\STATE Using the selected $\tau$, $\alpha_c$, and $\lambda$, compute $R^{\text{rsma}}_{s}$ and $R^{\text{rsma}}_{w}$ as in (\ref{7}) and (\ref{8}), respectively.
\end{algorithmic}
\end{algorithm}
\begin{figure}[t] \vspace{-0.1in} 
     \centering
     \captionsetup{justification=raggedright,singlelinecheck=false}
     \includegraphics[width=8cm,height=10cm,keepaspectratio]{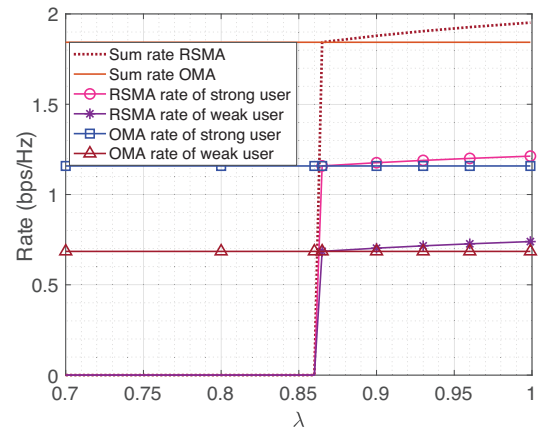}
     \vspace{-0.1in}
     \caption{Variation of the rates of $s$ and $w$ w.r.t $\lambda$.}
     \label{fig:fig4} \vspace{-0.2in}
\end{figure}
\begin{figure}[t]
     \centering
     \captionsetup{justification=raggedright,singlelinecheck=false}
     \includegraphics[width=8cm,height=10cm,keepaspectratio]{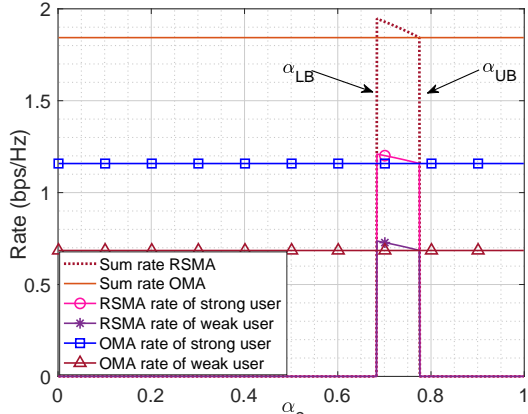}
     \vspace{-0.13in}
     \caption{Variation of the rates of $s$ and $w$ w.r.t $\alpha_c$.}
     \label{fig:fig5} \vspace{-0.2in}
\end{figure}
\vspace{-0.2in} 
\section{Results and Discussions}\label{results}
In this section, we validate the aforementioned bounds through simulation. We consider a two users scenario with interference as in \cite{l8}. Fig. \ref{fig:fig4} illustrates the variation of RSMA and OMA rates (in bps/Hz) of $s$ and $w$ w.r.t. $\lambda$. For $\gamma_s = 6$ dB, $\gamma_w = 2$ dB, $\beta = 0$ (i.e., for perfect SIC), $P_t = 1$ W, we have observed that with increase in $\lambda$, the $SR^{\text{rsma}}$ increases beyond $SR^{\text{oma}}$ after $\lambda = 0.865$. The $\tau$ values computed using $\lambda < 0.865$ are negative, which is inappropriate for computing the rates. Hence, the RSMA rates in the region, $\lambda < 0.865$, are indicated as zero. The value of $\lambda = 0.865$ is greater than the value derived using (\ref{lam1}), (\ref{lam2}), (\ref{lam4}), and (\ref{lam5}). This shows that the bound obtained using (\ref{lam1}), (\ref{lam2}), (\ref{lam4}), and (\ref{lam5}) is not strict. Thus, the lower bound of $\lambda$ computed using (\ref{pc2_2}) and presented in Fig. \ref{fig:fig2} is validated in Fig. \ref{fig:fig4}. Nevertheless, selecting the $\lambda$ using (\ref{lam1}), (\ref{lam2}), (\ref{lam4}), and (\ref{lam5}) is necessary to make the bound of $\alpha_c > 0$. Furthermore, it is observed from Fig. \ref{fig:fig3} for $\lambda > \lambda_{\text{lower}}$ that $R^{\text{rsma}}_{s}$ and $R^{\text{rsma}}_{w}$ are greater than $R^{\text{oma}}_{s}$ and $R^{\text{oma}}_{w}$, respectively. Note that the private message of user $s$ requires more power than that of user $w$.

Fig. \ref{fig:fig5} shows the variation of rate w.r.t $\alpha_c$ for $\beta=0$ for the same set of $\gamma_s$ and $\gamma_w$. For $\lambda = 0.99$, the lower and upper bound of $\alpha_c$ computed using (\ref{pc_lower}) and (\ref{pc2_2}), respectively, are $0.683$ and $0.776$. It is evident from the Fig. \ref{fig:fig5} that for $\alpha_{\text{LB}} < \alpha_c < \alpha_{\text{UB}}$, the individual RSMA rates of $s$ and $w$ are greater than their respective OMA rates. Moreover, the lower and upper bound of $\alpha_c$ computed using (\ref{pc_lower}) and (\ref{pc2_2}) are approximately equal to the $\alpha_{\text{LB}}$ and $\alpha_{\text{UB}}$ observed in Fig. \ref{fig:fig5}. Furthermore, similar results are obtained by varying $\tau$ for $\lambda = 0.99$ and $\alpha_c = 0.689$. However, due to space constraints, we cannot present the plots.
\begin{figure}[t]
     \centering
     \captionsetup{justification=raggedright,singlelinecheck=false}
     \includegraphics[width=8cm,height=10cm,keepaspectratio]{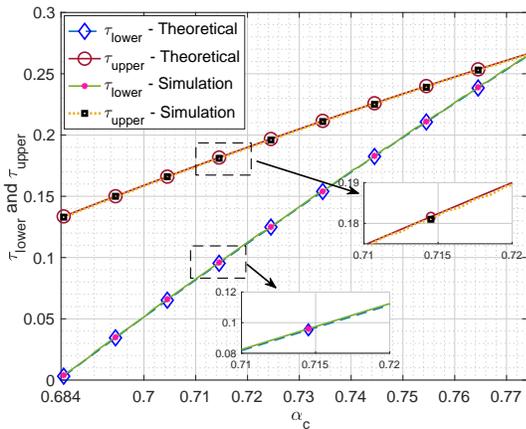}
     \vspace{-0.1in}
     \caption{Variation of $\tau_{\text{lower}}$ and $\tau_{\text{upper}}$ w.r.t $\alpha_c$ for $\lambda = 0.99$.}
     \label{fig:fig6} \vspace{-0.2in}
\end{figure}
\begin{figure}[t]
     \centering
     \captionsetup{justification=raggedright,singlelinecheck=false}
     \includegraphics[width=8cm,height=10cm,keepaspectratio]{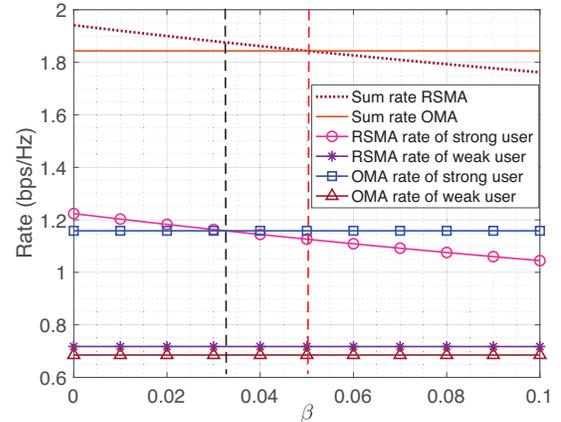}
     \vspace{-0.1in}
     \caption{Variation of the rates of $s$ and $w$ w.r.t $\beta$.}
     \label{fig:fig7} \vspace{-0.2in}
\end{figure}

Since the derived bounds of $\tau$ as shown in (\ref{tau_lb}) and (\ref{tau_ub}) are functions of  $\alpha_c$ and $\lambda$, we have plotted the variation of the bounds w.r.t $\alpha_c$ for $\lambda = 0.99$ in Fig. \ref{fig:fig6}. $\alpha_{\text{LB}}$ and $\alpha_{\text{UB}}$ are computed using (\ref{pc_lower}) and (\ref{pc2_2}), respectively. For $\beta = 0$, $\gamma_s = 6$ dB, and $\gamma_w = 2$ dB, $\tau_{\text{lower}}$ and $\tau_{\text{upper}}$ are plotted by varying $\alpha_c$ between $\alpha_{\text{LB}}$ and $\alpha_{\text{UB}}$. We can observe that these bounds are almost equal to the theoretical bounds computed using (\ref{tau_lb}) and (\ref{tau_ub}). Similar results can be obtained for different $\lambda$. 
% For the same set of $\gamma_s$ and $\gamma_w$, $\lambda = 0.99$, and $\alpha_c = 0.689$, the $\tau_{\text{lower}}$ and $\tau_{\text{upper}}$ computed using (\ref{tau_lb}) and (\ref{tau_ub}) are approximately $0.0173$ and $0.1410$, respectively. Using these parameters, when we vary $\tau$ in the [0,1] and compute the rate, the $\tau_{\text{lower}}$ and $\tau_{\text{upper}}$ observed in Fig. \ref{fig:fig6} are approximately equal to the bounds derived using (\ref{tau_lb}) and (\ref{tau_ub}). We can observe that only in that region, the RSMA rates of $s$ and $w$ are greater than their respective OMA rates.

Fig. \ref{fig:fig7} shows the variation of rate w.r.t. $\beta$. We can observe that as $\beta$ increases, after 0.035, the $R^{\text{rsma}}_{s}$ becomes less than $R^{\text{oma}}_{s}$. However, the $SR^{\text{rsma}}$ remains greater than $SR^{\text{oma}}$ as $\beta$ increases further. Nevertheless, after $ \beta = 0.05$, $SR^{\text{rsma}}$ starts deteriorating. This is due to the decrease in $\gamma_{ps}$ and $\gamma_{pw}$ with an increase in $\beta$. The result in Fig. \ref{fig:fig7} is plotted for $\lambda = 0.99$, $\alpha_c = 0.689$, and $\tau = 0.1$.\vspace{-0.1in}
\section{Conclusion}\label{conclusion}
In this letter, we have derived the bounds on the power allocation coefficients of the common and private messages and the fraction of the individual common message of the users for which the RSMA rates of users will be greater than their respective OMA rates. We have validated the derived bounds using simulation. We have numerically computed $\alpha_c$, $\lambda$, and $\tau$, which are validated using simulation. In our future work, we aim to compute the closed-form expressions of the lower bound of $\lambda$ and $\alpha_{\text{UB}}$. We also aim to derive bounds on $\beta$ in our future work.\vspace{-0.2in}
% \section{Acknowledgement}
% This work was supported in part by the SERB NPDF--project: PDF/2020/001251, Indo-Norwegian Collaboration in Autonomous Cyber-Physical Systems (INCAPS)--project: 287918 of the INTPART program, the Low-Altitude UAV Communication and Tracking (LUCAT)--project: 280835 of the IKTPLUSS program from the Research Council of Norway and the Department of Science and Technology (DST), Govt. of India, and DST NMICPS through TiHAN Faculty fellowship of Dr. Abhinav Kumar.
\bibliography{ref}
\bibliographystyle{IEEEtran}
\vspace{14mm}
\end{document}